\documentclass[letterpaper]{article} 
\usepackage{aaai2026}  
\usepackage{times}  
\usepackage{helvet}  
\usepackage{courier}  
\usepackage[hyphens]{url}  
\usepackage{graphicx} 
\usepackage{natbib}  
\usepackage{caption} 
\usepackage{algorithm}
\usepackage{algorithmic}

\usepackage{newfloat}
\usepackage{listings}
\DeclareCaptionStyle{ruled}{labelfont=normalfont,labelsep=colon,strut=off} 
\floatstyle{ruled}
\newfloat{listing}{tb}{lst}{}
\floatname{listing}{Listing}
\usepackage{dirtytalk} 
\usepackage{enumitem} 
\usepackage{subcaption, caption, multirow} 

\title{Biologically-Informed Hybrid Membership Inference Attacks on Generative Genomic Models}
\author {
    Asia Belfiore\textsuperscript{\rm 1},
    Jonathan Passerat-Palmbach\textsuperscript{\rm 1},
    Dmitrii Usynin\textsuperscript{\rm 2}
}
\affiliations {
    \textsuperscript{\rm 1}Imperial College London\\
    \textsuperscript{\rm 2}Technical University of Munich\\
    belfiore.asia.02@gmail.com, j.passerat-palmbach@imperial.ac.uk, dmitrii.usynin@tum.de
}

\begin{document}

\maketitle

\begin{abstract}
  The increased availability of genetic data has transformed genomics research, but raised many privacy concerns regarding its handling due to its sensitive nature. This work explores the use of language models (LMs) for the generation of synthetic genetic mutation profiles, leveraging differential privacy (DP) for the protection of sensitive genetic data. We empirically evaluate the privacy guarantees of our DP modes by introducing a novel Biologically-Informed \textbf{Hybrid Membership Inference Attack} (biHMIA), which combines traditional black box MIA with contextual genomics metrics for enhanced attack power. Our experiments show that both small and large transformer GPT-like models are viable synthetic variant generators for \textit{small-scale} genomics, and that our hybrid attack leads, on average, to higher adversarial success compared to traditional metric-based MIAs. 
\end{abstract}

\section{Introduction}
 
Genomics is the study of the structure and function of an organism's genetic information: all humans share over 99\% of their DNA, and less than 1\% of the entire genetic information dictates the unique variety of human observable traits. Differences between individuals' DNA, known as \textit{genetic mutations}, most commonly comprise of changes (deletions, insertions and substitutions, or \textit{Single Nucleotide Polymorphisms} (SNP)) of nucleic bases called \textit{alleles}, in specific genomic location (or \textit{loci}) compared to the reference human genome. The \say{rarity} of a specific allele is measured via \textit{Allele Frequency} (AF) or \textit{Minor Allele Frequency} (MAF), as the prevalence of the rarest occurring base within all alleles in the study cohort, with mutations being classified as rare if they appear in less than 0.5-1\% of the population. 
In the early 2000s, the efforts of the Human Genome Project (HGP) \cite{collins_human_2003} led to the publishing of the first \textit{sequenced} human genome, and represented the very beginning of the Genomic Era. In the following years, as sequencing costs decreased due to the development of high-throughput technologies and tools \cite{gauthier_brief_2019}, the amount of available genetic data saw an exorbitant increase, pushing the birth of major public international bioinformatics initiatives like the 1000 Genomes Project \cite{the_1000_genomes_project_consortium_global_2015} (1000GP), and leading to breakthrough advancements in personalised healthcare and disease discovery \cite{goyal_synthetic_2023}. Such steep growth of publicly shared genetic data raised concerns regarding the safety of its handling and storage \cite{gauthier_brief_2019}, especially due to the lack of unified regulations around Genomic Privacy \cite{shen_privacy_2017}. Furthermore, emerging threats in genomics are highlighting the need of clear methodology for protecting  genomic owners' privacy \cite{shen_privacy_2017}.

\subsection{Synthetic Data} 

Genomic data is particularly sensitive, as it holds an individual's \say{most intimate kind of information} \cite{bains_genetic_2010} and possesses unique traits that distinguish it from others, including providing information about more than just the individuals themselves \cite{naveed_privacy_2015}. Due to this, the risks linked to the leakage of genomic data pose serious legal, ethical and social concerns, exacerbated by the \say{unknown future risks} that may arise as the genomics and bioinformatics fields evolve \cite{naveed_privacy_2015}. The use of \textit{Synthetic Data}, i.e. information which mimics the nature, format and statistical properties and distributions of some real data without exposing it \cite{information_commissioners_office_uk_2025}, could have a significant and beneficial impact in genomics. Not only would it enable researchers to create ad-hoc datasets tailored for their study scope, but it could mitigate the extensive privacy issues and concerns around the handling of genomic data, thanks to the (ideally) total absence of references to the original information \cite{guepin_synthetic_2023}, allowing scientists to freely and safely share the generated datasets for the benefit of the scientific community. 

\subsection{Generative Models} 

A promising solution to the issue of public data scarcity or restrictions in sectors that deal with highly sensitive data, like genomics, lies in the exploitation of generative models, like \textit{Language Models} (LMs), to create new synthetic data points. Statistically speaking, a \textit{generative model} aims to reproduce some conditional probability distribution $P(X|Y=y)$ based on some input and output variables X and Y, in order to generate new instances $x$ of the observable variable $X$ \cite{gm_comprehensive_2020}. Machine Learning (ML) based generative models have so far found many applications in healthcare \cite{sen_opportunities_2024}, as their ability to incorporate non-linear patterns found within the data makes them especially useful for modelling complex data types like health-related information \cite{goyal_synthetic_2023}. Today's LLMs have proven to be suitable for the modelling of a wide variety of data, including synthetic health records \cite{huang_pangenome-informed_2024, doi:10.1126/science.aaa8685} and sequencing genetic data \cite{huang_pangenome-informed_2024}. Due to the sensitive nature of genomics data, it is imperative to ensure that the trained models are adequately secure for public sharing, as they may leak at-risk information about individuals in the dataset like presence of rare mutations or diseases, via \textit{privacy attacks} like Membership Inference Attacks.

\subsection{Membership Inference Attacks} 

Membership Inference Attacks (MIA) aim to infer the presence of a target individual's data within a given dataset, and were first introduced as attacks to ML models in order to infer whether or not a target record is part of the model's training dataset \cite{shokri_membership_2017, rigaki_survey_2024}. Protection against MIAs revolves around hiding the presence of individuals by making it impossible to distinguish between a dataset that includes some target record, and one that doesn't. This desirable \say{ambiguity} is exactly provided and mathematically guaranteed by \textit{Differential Privacy} (DP), a mechanism that injects noise into the data in order to ensure nothing can be learned about a target individual from the dataset itself \cite{dwork_exposed_2017}.
MIAs represent the current standard method for the assessment of a model's privacy robustness. Today, many highly successful MIAs have been taken out on ML models with regards to various types of data, including healthcare and genomics \cite{shokri_membership_2017, hu_membership_2021, salem_ml-leaks_2018, homer_resolving_2008}. However, traditional sample-level MIA may be unsuitable in the context of synthetic genomics, due to the nature of the data itself.

\section{Privacy in Genomics} \label{sec:hybrid_mia}

Traditional anonymization techniques have been shown to be inadequate for the protection of genomic owners' privacy, as anonymised genomes in public databases (like the 1000GP) have been frequently used to carry out re-identification and Inference Attacks \cite{naveed_privacy_2015}.
In fact, although on average a human genome will contain tens of millions of mutations, studies \cite{thomas_assessing_2024, shabani_reidentifiability_2019} have shown that variant combinations of as little as 30 to 80 SNPs can uniquely identify an individual, and thus their presence in an anonymized cohort can be easily inferred by gaining access to a small part of an individual's genome. Similarly, a study \cite{sankararaman_genomic_2009} found that it is possible to infer the genotypes of participants in pooled studies by simply having access to a reduced number of so-called \say{\textit{exposed SNPs}}. Samani et. al. \cite{samani_quantifying_2015} further showed that it is possible to uncover an individual's missing SNPs from their partially known genotype even when up to 40\% of their genomic information is {hidden}. Homer et. al. \cite{homer_resolving_2008} were able to re-identify individual participants in genomic mixture data based on genotypes with a high-SNP density, while another study \cite{schadt_bayesian_2012} was able to identify individuals in large cohorts by statistically predicting their variant genotypes based on known SNPs and gene expression data. \\
Furthermore, the expanding use of ML in genomics has created a whole new frontier of risks beyond the raw genetic information itself, as manipulation or simple inference of the trained models can potentially lead to genetic privacy breaches of the underlying training data. LMs, especially large-scale ones with billions of trainable parameters, generally tend to exhibit high levels of \textit{memorization}, i.e. re-generation of exact data sequences seen during the training or fine-tuning process, which raises significant issues when the generative task involves sensitive data \cite{carlini2021extractingtrainingdatalarge} (like personal or genetic data), as it could lead to exposure of highly identifiable information of samples in the training dataset.

\subsection{Biologically-Informed Hybrid Membership Inference} 

Most individuals share a large portion of their sequences and mutations with the majority of the population, by nature of genomics data itself. Traditional MIAs assume the presence of unique sets of features for each sample, and thus may fail in the genomics context as it may be unable to appropriately capture the uniqueness of each individual. Due to such lack of a clear definition of a successful MIA tailored specifically on genetic data, we hypothesise that introducing genomics-specific context into the feature set may yield a stronger attack. We argue that adding genomic-specific information into the feature set used for membership inference may reveal and capture subtle patterns and nuances between both the model's \say{passive behaviour} (captured via model metrics, also used in traditional MIA) and \say{active behaviour} (captured via extracted features on the model's generative output) to seen versus unseen sequences, and better capture unique patterns of the individuals in the dataset.  \\
We employ a feature-extraction framework to create a \textit{hybrid feature set} based on a mix of model-based and genomics-based metrics to use for membership inference. We pass original sample profiles to the trained model's forward pass to extract model logits and losses from the target input sequences. We then prompt the model with some starting mutation from the original target sample to obtain a synthetic profile for each of the individuals in the target cohort, and use the generated profiles to extract the \say{genomic-specific} features. Based on metrics and statistics commonly extracted from genomic mutation data \cite{10.1093/bioinformatics/btr330}, we chose the final feature set used for the attack to comprise of:

\begin{itemize}
    \item \textbf{Model-based features}: \textit{Perplexity}, \textit{Average Confidence}, \textit{Average Loss} and \textit{Loss Variance}, and \textit{Average Logit Magnitude}, calculated from the model logits and loss on the given input sequences (Table~\ref{tab:model-features}).
    \item \textbf{Domain-specific features}: \textit{Mutation Rates} (ratio of mutated variants over all generated sample mutations), \textit{Genotype Frequencies} (frequencies of genotypes),  \textit{Variation Frequencies by type} (number of deletions, insertions and substitutions) and \textit{by nature} (number of biallelic and multiallelic variants), extracted from the \textit{synthetic generated} sequence (Table~\ref{tab:genomic-features}). 
\end{itemize}

\section{Methods}

\subsection{Data and Preprocessing} 

We use individual-level \textit{Variant Call Format} (VCF) data of 2504 samples from the 1000 Genomes Project Phase 3 files \cite{the_1000_genomes_project_consortium_global_2015}. We focus on genetic mutations present on Chromosome 22 for all study samples in order to reduce computational overheads given its lower density of mutations, all while preserving the format and statistics of the data. To best allow the model to understand the underlying variants distributions in a sample-wise manner, we extracted individual genetic profiles from the source VCF file by querying the VCF dataset variant-wise (row-wise) and encoding each mutation as a string '$CHR:POS:REFs>ALTs\_GT$' to introduce some context into the input, while keeping the amount of non-genomics related tokens to a minimum. 

\subsection{Models} 

We focused on generative models for \textit{Causal Language Modelling} (CLM), as they learn to predict the next mutation in the sample profile based on the previously seen mutations, which we believed would allow for better contextualization and generation of biologically-plausible data. We finetuned OpenAI's GPT-2 Model \cite{radford2019language} and trained an adaptation of \textit{MinGPT} \cite{Karpathy-minGPT}, a small, fast and interpretable reimplementation of the GPT-2 transformer that allows for full parameter and training customization.
We chose model parameters and training regimes under following hypotheses: (\underline{H1}) \textbf{Smaller models provide better privacy}: we fine-tuned the smallest version of GPT-2 (124M parameters), and made MinGPT significantly smaller ($\sim$ 12M parameters) than the pretrained counterpart, under the hypothesis that smaller models may provide better privacy compared to larger ones by being less prone to memorization; (\underline{H2}) \textbf{Training from-scratch enhances utility}: training a generative model from scratch on the custom genomic dataset could lead to better utility compared to finetuning a general-use model.
For MinGPT, we implemented a custom Regex-based BPE Tokenizer \cite{Karpathy-regex}, which pre-tokenizes the corpus following a custom \textit{Regex} pattern created from the format of the training mutations, retaining the biologically-appropriate structures of the data by \say{chunking} the corpus into individual mutations.
We introduced DP mechanisms during the model training via Differentially Private Stochastic Gradient Descent (DP-SGD) \cite{abadi_deep_2016}, which limits the privacy loss after every gradient update by performing \textit{noise addition} and \textit{norm clipping} (scaling the gradient based on some pre-defined threshold to bound the influence of each sample). In the following sections, we refer to the DP-trained versions of our models by suffixing them with -DP.

\subsection{Utility and Privacy Evaluation} 

The utility evaluation assesses the ability of the model to generate data that follows the same format and statistics of the training data. We evaluate the general quality of each generated sample in terms of \textit{Validity} (ratio of mutations generated in the valid format) and \textit{Quality} (ratio of mutations generated within biologically valid positional bounds for the chosen chromosome). Furthermore, we evaluate the cohort-level quality and diversity of the generated data via \textit{Uniqueness} vs \textit{Repetition} (ratio of mutations generated for only one vs at least two samples in the synthetic cohort) and \textit{Novelty} vs \textit{Memorization} (ratio of mutations unseen vs seen during training). We also extract \textit{VCF and Variant statistics} including frequency of variants by type (biallelic and multiallelic) and nature (deletions, insertions and substitutions) and mutation-wise sample and allele statistics. In-depth description of these metrics can be found in the Appendix Table \ref{tab:vcf-stats}, and extracted the same metrics from the original training data to use as benchmarks. \\
The privacy evaluation, instead, assesses to which extent (if any) the model tends to \textit{leak} information about the training data, thus defining the \textit{privacy risks} of using and sharing the developed models.
We use the aforementioned set of hybrid features to carry out two types of MIA attacks: a \textit{Threshold-based} attack, to distinguish between training and holdout targets based on the respective density of data points which fall within a pre-defined threshold, and \textit{ML-based}, which involves training classifiers (K-Nearest Neighbour, Logistic Regression and Random Forest) to distinguish between features extracted from the training data and from the holdout data. 

\section{Results}

\subsection{Utility Evaluation} 

In general, all tested models proved to be viable synthetic genomic sample generators, producing correctly formatted data and (mostly) positionally valid mutations. When compared to the validity benchmark, the quality scores from the generative models are satisfactory, and empower us to confirm that these models can be employed, at least, for mock sample mutation profiles generation. In terms of advanced statistics, MinGPT-DP showed best simulation power in terms of mutation type, nature and variant distribution, but none of the implemented models successfully reproduced the original VCF statistics, with notably scarce abilities to mimic the chromosomal mutation distributions and frequent generation of variants outside the real chromosomal ranges.

\subsection{Privacy Evaluation} 

\begin{figure}[H]
    \centering    
    \begin{subfigure}{\linewidth}
          \centering
          \includegraphics[width=\linewidth]{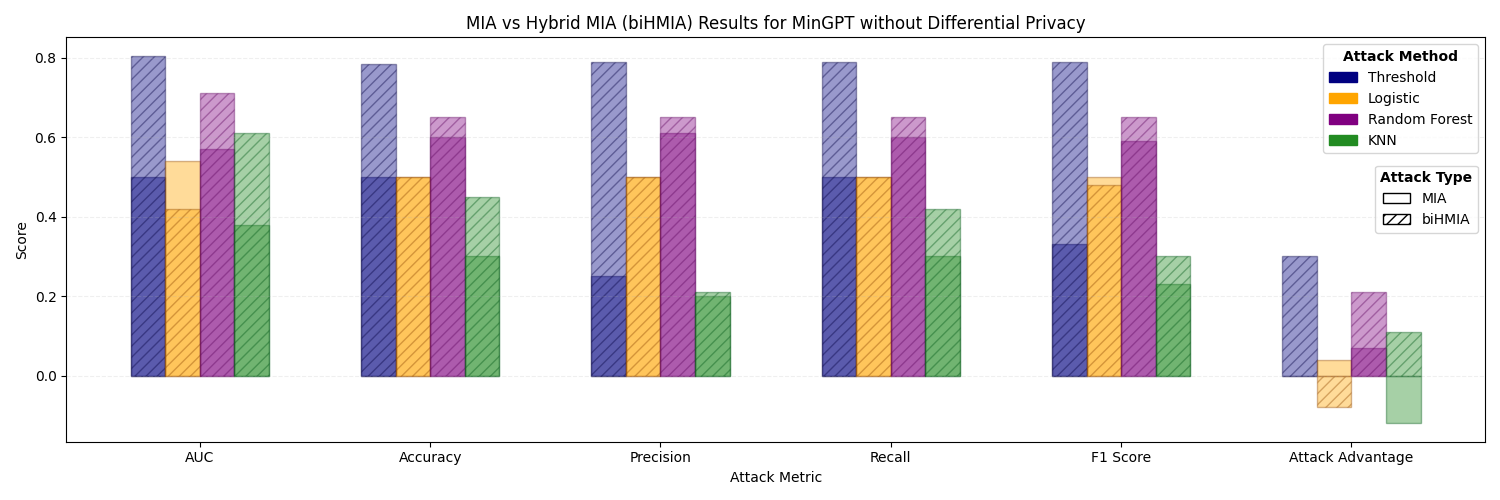}
          \caption{MIA vs Hybrid MIA (biHMIA) scores on MinGPT without DP.}
        \end{subfigure}
        \hfill
        \begin{subfigure}{\linewidth}
          \centering 
          \includegraphics[width=\linewidth]{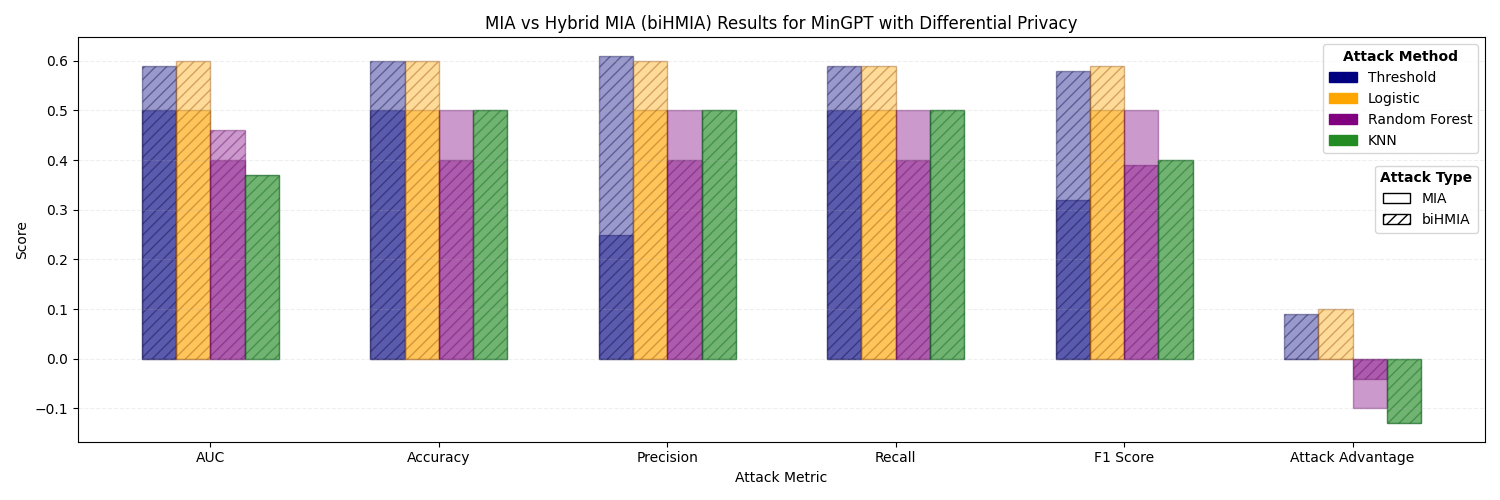}
          \caption{MIA vs Hybrid MIA (biHMIA) scores on MinGPT with DP $\epsilon=1$.}  
        \end{subfigure}
    \caption{Comparison of Model-Based versus Hybrid MIA on MinGPT trained without (a) and with DP ($\epsilon=1$) (b). It shows from left to right: AUC, Accuracy, Precision, Recall, F1-Score and Attack Advantage for Threshold Attack, Logistic Regression, Random Forest and K-Nearest Neighbour.}
    \label{fig:MIA-results}
\end{figure}

We found the smaller models to naturally offer more robust privacy guarantees compared to the finetuned models, which instead showed significant levels of memorization. The smaller models had, on average, lower MIA scores compared to their GPT-2 counterparts (Figures \ref{fig:MIA-results} and \ref{fig:MIA-GPT2-results}). These results confirm our hypothesis (\underline{H1}) that smaller models may be a better fit for highly-sensitive generation tasks, as they tend to have lower memorization capacity and a higher level of stochasticity.
We generally found a decrease in MIA scores between non-DP and DP-trained models, which we expected.
We found that our hybrid MIAs have generally higher inference success, especially on the non-DP models, and thus represent a promising novel approach for the evaluation of genomics-specific privacy guarantees of generative models.

\section{Discussion} 

\subsection{Innate Privacy and Differential Privacy Effects} 

We speculate that the noise addition introduced by DP acts as a regularizer for the model, reducing model hallucination, which we speculate may be responsible for the positional validity and quality increase in the DP-trained MinGPT-generated sequences. To this we also attribute the increase in mutation uniqueness in the DP-trained MinGPT, speculating that that the noise addition increases the sampling randomness during the generation process leading to more stochastic generation tendencies.

\subsection{Results Variance} 

All attacks were subject to significant fluctuation between rounds on different randomly-chosen target cohorts. We experienced a non-insignificant level of incongruity between different hybrid MIA runs on the same data, particularly for the MinGPT models. We speculate that this is due to variations within samples generated on different runs (and, thus, variability among the extracted genomic features), which we attributed to the higher model stochasticity. This is further supported by the fact that we did not find the same high variability for GPT-2, which tended to produce more repetitive samples. We also found that running the same attacks on a smaller target cohort generally led to higher MIA success compared to attacks taken out against a larger number of target samples, as we speculate that the classifiers may struggle to generalize on larger cohorts as they may include a higher number of member and non-member samples that share very similar features, making it harder to differentiate between them.

\subsection{Limitations} 

While overall the results we obtained were promising (both utility and privacy wise), we must highlight that the limited size and diversity of the dataset used in these experiments may have biased our assessments. Following this, we hypothesise that the smaller models outperform the larger ones due to simply being better suited for such modest training setup, as the larger models require much more broad and extensive datasets. From the obtained results, we deduce the larger model to have \textit{overfit} on the training data, evidenced by the very quick training loss convergence and high levels of memorization and repetition observed for the GPT-2 model.

\section{Conclusion}

Our research explored the suitability of Language Models for the generation of synthetic sample-level genomic mutation profiles, leveraging Differential Privacy to enhance the models' privacy guarantees. We introduce a hybrid MIA attack which combines model metrics with genomic-specific features extracted from the generated samples to infer membership. This attack shows on average higher success rates compared to model-only based MIA, thus representing a promising solution for genomics-specific privacy assessments of generative models. We found that GPT-like transformer-based models are valid generators of synthetic genomic samples, both if trained from scratch and if finetuned. Smaller models tend to offer more inter-sample diversity within generated multi-sample cohorts, but at the cost of positional verisimilitude. We found that smaller transformers provide better innate privacy, with low memorization and good robustness against MIAs. Larger models, on the other hand, are more vulnerable to MIAs and tend to show high levels of memorization. Furthermore, implementing DP into the training pipeline increases the privacy robustness of the model to inference attacks, and produces a regularization effect that enhances the model’s utility.

\subsection{Future Directions} 

Training or finetuning the models on larger more diverse datasets, like using mutations from \textit{all} chromosomes or leveraging data from large biobanks \cite{10.1371/journal.pmed.1001779}, could lead to better utility. We also believe that evaluating the utility and privacy guarantees of increasingly larger version of MinGPT could unveil the relationship between model utility/privacy and parameter size, leading to some \say{middle ground model} that offers a natural balance between utility and privacy. Implementing contextual information within the training samples, like population or sex, could further enhance the models' utility by adding contextual dependencies that would allow the models to better understand the data properties.

\bigskip
\bibliography{references}

\appendix

\section{Appendix}

\section{Supplementary Information}

\subsection{Utility Metrics}

We chose the metrics described below as measures of synthetic data quality, diversity and genomic credibility. We use these to assess whether the developed model is suitable for the generation of the desired data based on benchmarking values extrapolated from the original training data. In particular, we define the generative model to be satisfactory for \textit{mock data generation} if it provides individual- and cohort-level mutational quality, validity and diversity scores comparable to the original data (Table~\ref{tab:utility-stats}).

\begin{table}[H]
    \setlength{\tabcolsep}{6pt}
    \renewcommand{\arraystretch}{1.5}
    \centering
    \begin{tabular}{p{2cm}|p{5.5cm}}
        \hline
        {Utility Metric} & {Description} \\
        \hline
        Validity & Represents the ratio of mutations generated in the valid format out of all generated mutations. \\
        \hline
        Quality & Represents the ratio of mutations generated within biologically valid positional bounds for the chosen chromosome, out of all generated mutations. \\
        \hline
        Uniqueness & Represents the ratio of mutations uniquely generated for a single sample. \\
        \hline
        Repetition & Represents the ratio of mutations generated for at least two samples out of all generated synthetic individuals. \\
        \hline
        Novelty & Represents the ratio of novel/unseen mutations amongst all the generated synthetic variants. \\
        \hline
        Memorization & Represents the ratio of mutations memorized from the training samples amongst all generated synthetic variants. \\
        \hline
    \end{tabular}
    \caption{Minimal metrics used for mock data utility evaluation.}
    \label{tab:utility-stats}
\end{table}

We define the model to be satisfactory for \textit{synthetic data generation} if it provides, on top of the aforementioned mock-data requirements, domain-specific statistics comparable to the original data (Table~\ref{tab:vcf-stats}). These represent distributions and patterns that characterize the source data, which determine the biological validity of the generated mutations.

\begin{table}[hp]
    \setlength{\tabcolsep}{6pt}
    \renewcommand{\arraystretch}{1.5}
    \centering
    \begin{tabular}{p{2cm}|p{5.5cm}}
        \hline
        {Utility Metric} & {Description} \\
        \hline
        Variant Statistics & Describe the frequency of variants by types (biallelic and multiallelic) and nature (deletions, insertions and substitutions). \\
        \hline
        \multirow{2}{4cm}{INFO fields} & Describe the data in terms of sample and allele statistics, calculated mutation-wise. In particular, for a given mutation: \\
        \cline{2-2}
        & – NS: Number of samples that have the mutation (any genotype).\\
        \cline{2-2}
        & – AC: Allele count amongst all genotypes for the mutation, for each alternate allele. \\
        \cline{2-2}
        & – AF: Allele frequency for each present allele across all samples. \\
        \cline{2-2}
        & – AN: Total number of alleles in the samples genotypes. \\
        \cline{2-2}
        & – aaf: Allele frequency for each present alternate allele. \\
        \hline
        FILTER fields & Represent the number of samples with data for that mutation, out of the total number of samples. Values in the form sX, indicate that less than X\% of samples “have data” for that mutation, for any number X. \\
        \hline
        Call Rate & Represents how many mutations are “called” out of all the available ones, i.e. the ratio of ’PASS’ values in the VCF FORMAT field, out of all mutations. \\
        \hline
        Genotypes & Describes the frequency of each genotype (0, 1, or more copies of the mutated allele) amongst all mutations. \\
        \hline
    \end{tabular}
    \caption{Advanced VCF-specific metrics used for synthetic data utility evaluation.}
    \label{tab:vcf-stats}
\end{table}

\subsection{biHMIA Features}

To carry out both the model-based MIA and the Hybrid biHMIA, we pass the tokenized original sample profiles to the model's forward pass by breaking them into \textit{subsequences} conforming to the model’s maximum input length, to obtain logits and losses used to extract the features described in Table~\ref{tab:model-features}.

\begin{table}[H]
    \setlength{\tabcolsep}{6pt}
    \renewcommand{\arraystretch}{1.5}
    \centering
    \begin{tabular}{p{2cm}|p{5.5cm}}
        \hline
        {Extracted Feature} & {Description} \\
        \hline
        Perplexity & Captures the level of model uncertainty on a given sequence, as the exponential of the sequence’s average negative log-likelihood. \\
        \hline
        Average Loss & The average model loss on a given sequence, calculated by averaging the losses over all input subsequences. \\
        \hline
        Loss Variance & Represents the average variance between losses of an input’s subsequences. \\
        \hline
        Average Logit Magnitude & Represents the average L2 norm across subsequence logits. \\
        \hline
        Average Confidence & Represents the average maximum softmax probability over all subsequences. \\
        \hline
    \end{tabular}
    \caption{Model-based features extracted from the input sequences for all (MIA and biHMIA) attacks.}
    \label{tab:model-features}
\end{table}

To extract the domain-specific feature set for the Hybrid biHMIA, we prompt the model with some starting mutation from the original target sequence to generate a synthetic profile for each of the target samples used for inference. We then extract the features listed in Table~\ref{tab:genomic-features} from the generated synthetic sequence to capture context on the model's generative process. We hypothesised that such genomic-focused features may better capture the biological uniqueness of each sample, and may reveal subtle biological patterns in the synthetic samples when generated from seen versus unseen mutations.

\begin{table}[H]
    \setlength{\tabcolsep}{6pt}
    \renewcommand{\arraystretch}{1.5}
    \centering
    \begin{tabular}{p{2cm}|p{5.5cm}}
        \hline
        {Extracted Feature} & {Description} \\
        \hline
        Mutation Rates & Represents the ratio of mutated variants over all generated sample mutations. \\
        \hline
        Genotype Frequencies & Represents the frequencies of homozygous reference, heterozygous reference and homozygous alternate genotypes for all generated sample mutations. \\
        \hline
        Variation Frequencies by Type & Represents the number of deletions, insertions and substitutions among all generated mutations. \\
        \hline
        Variation Frequencies by Nature & Represents the number of biallelic and multiallelic variants. \\
        \hline
    \end{tabular}
    \caption{Domain-specific features extracted from the synthetic generated sequences for the biHMIA attack.}
    \label{tab:genomic-features}
\end{table}

We evaluate the attacks' efficacy (both for MIA and biHMIA) by measuring each attack's Area Under the Curve (AUC), Accuracy, Precision, Recall, F-Score and Attack Advantage (as the difference between the attack’s AUC and random guessing, which quantifies the “information gain” yielded by the attack).

\section{Supplementary Figures}

\begin{figure}[H]
    \centering
    \includegraphics[width=\linewidth]{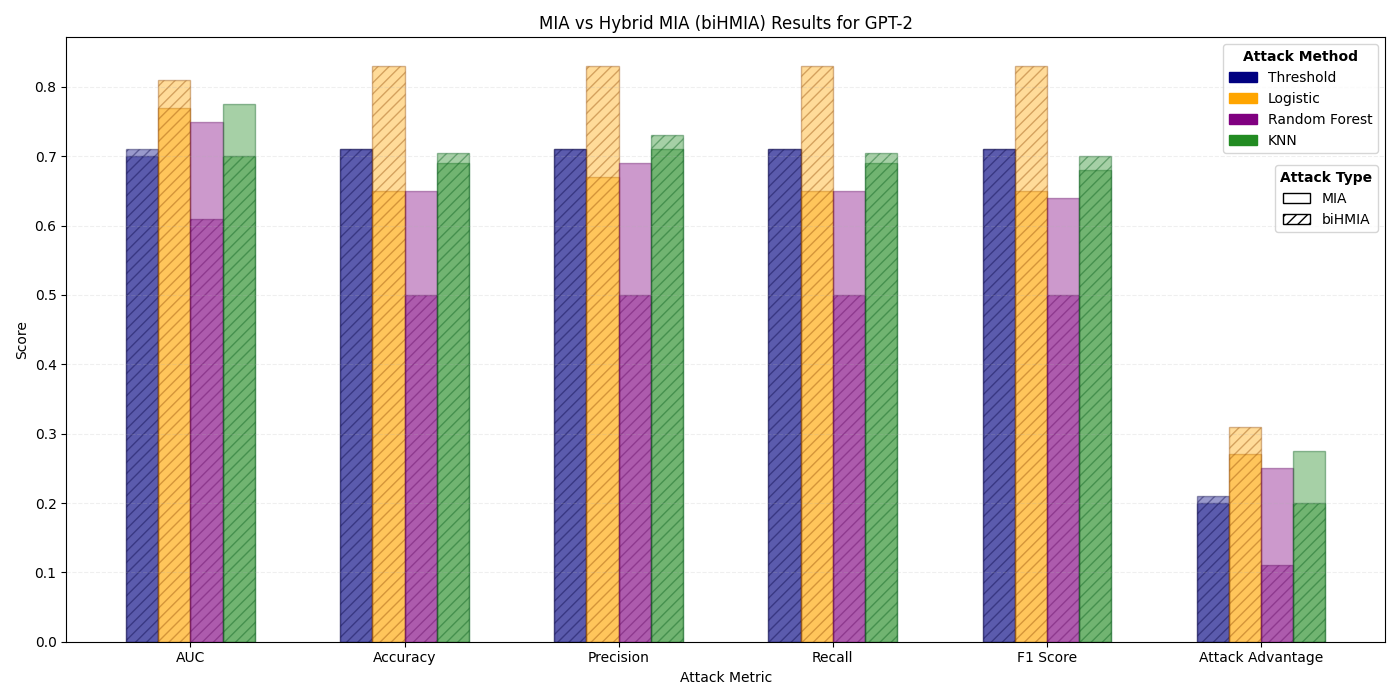}
    \caption{Comparison of Model-Based versus Hybrid MIA on GPT-2. It shows from left to right: AUC, Accuracy, Precision, Recall, F1-Score and Attack Advantage for Threshold Attack (in blue), Logistic Regression (in yellow), Random Forest (in purple) and K-Nearest Neighbour (in green).}
    \label{fig:MIA-GPT2-results}
\end{figure}

\begin{figure}[h]
    \centering    
    \begin{subfigure}{\linewidth}
          \centering
          \includegraphics[width=\linewidth]{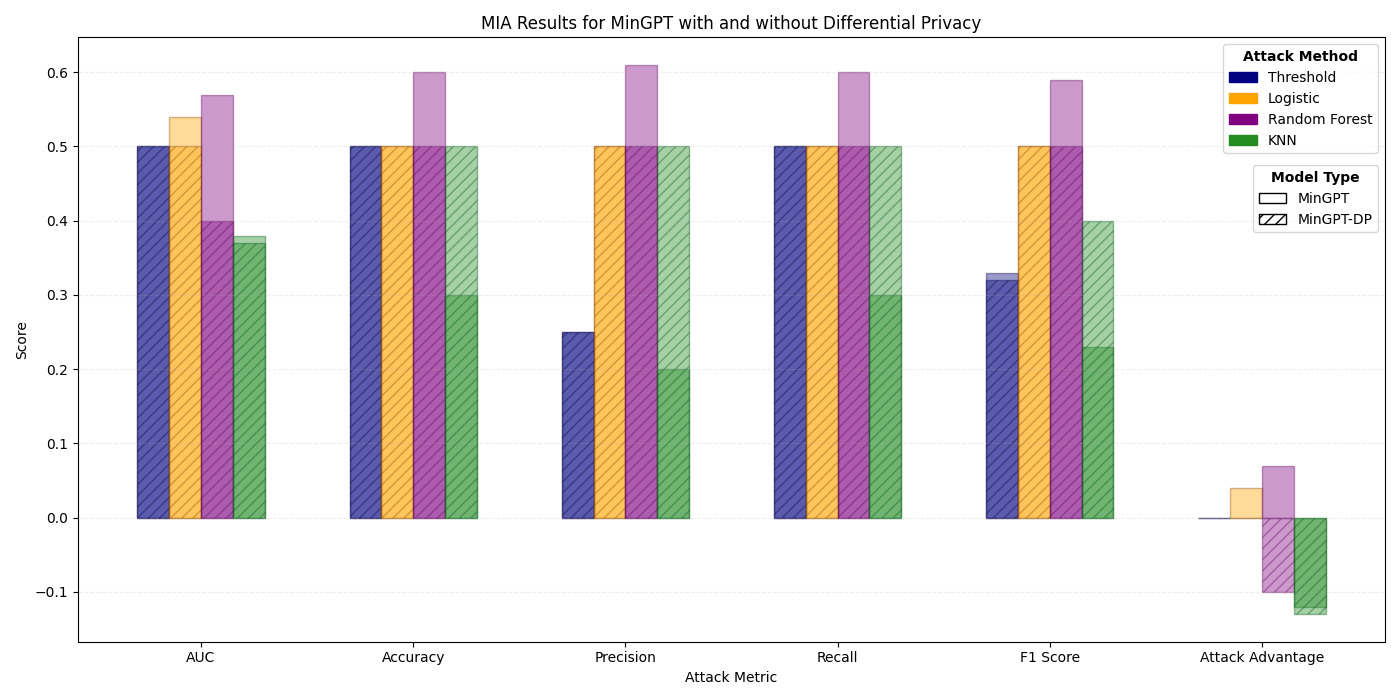}
          \caption{MIA scores on MinGPT (without and with DP $\epsilon=1$).}
        \end{subfigure}
        \hfill
        \begin{subfigure}{\linewidth}
          \centering 
          \includegraphics[width=\linewidth]{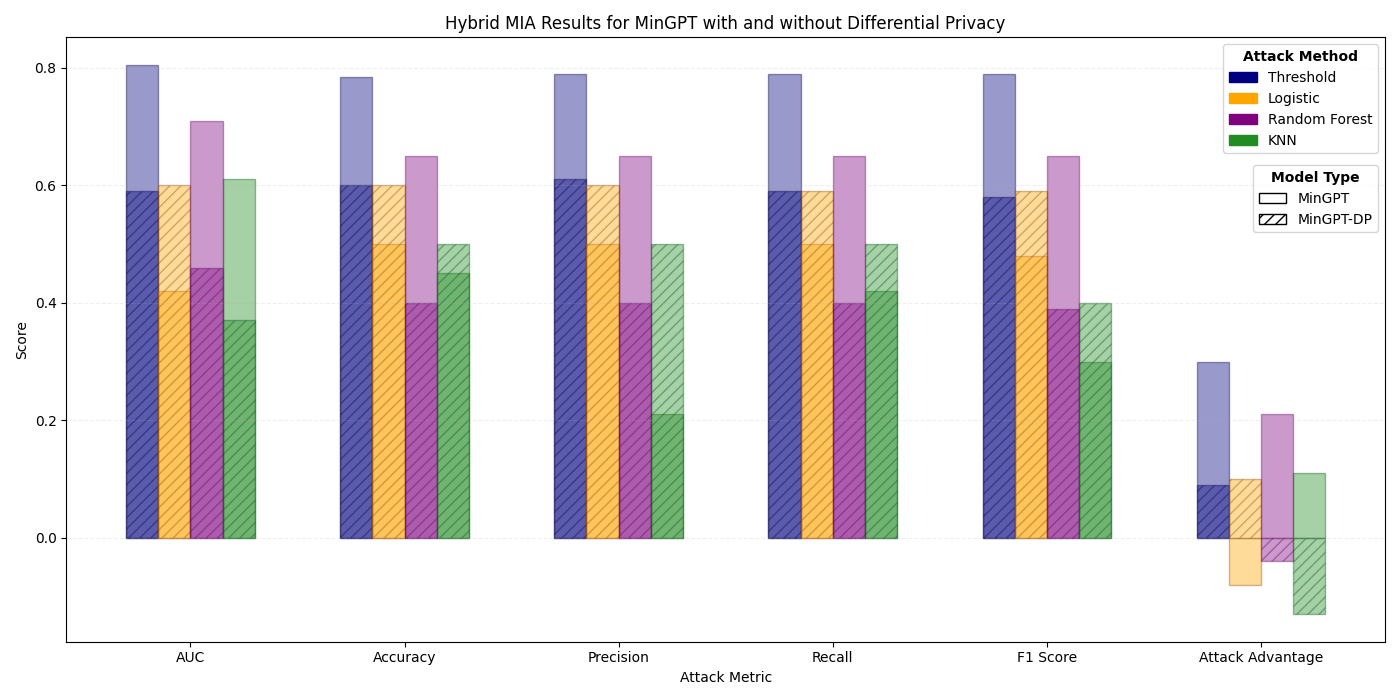}
          \caption{Hybrid MIA (biHMIA) scores on MinGPT (without and with DP $\epsilon=1$).}  
        \end{subfigure}
    \caption{Effects of DP ($\epsilon=1$) on Model-Based (a) versus Hybrid MIA (b) on MinGPT. It shows from left to right: AUC, Accuracy, Precision, Recall, F1-Score and Attack Advantage for Threshold Attack (in blue), Logistic Regression (in yellow), Random Forest (in purple) and K-Nearest Neighbour (in green).}
    \label{fig:mingpt_mias}
\end{figure}

\section{Supplementary Results}

In general, all evaluated models showed satisfactory ability to generate sample-level mutations profiles, with all DP and non-DP transformers achieving 100\% mutation quality (i.e. correctly formatted synthetic sequences).
We found the regularly finetuned GPT-2 model to achieve, in general, {slightly} better mutation quality scores in terms of mutation validity compared to the non-DP MinGPT. The DP-trained MinGPT achieves on average higher quality scores than its non-DP counterpart, with metrics comparable to the finetuned GPT-2 model. 
Interestingly, we found that MinGPT (both with and without DP) generates mutations outside the training chromosome (chromosome 22), including chromosomes 2, 8 and 16, suggesting some level of model hallucination. 

\begin{figure}[htp]
    \centering
    \begin{subfigure}[t]{0.49\linewidth}
          \centering
          \includegraphics[width=\linewidth]{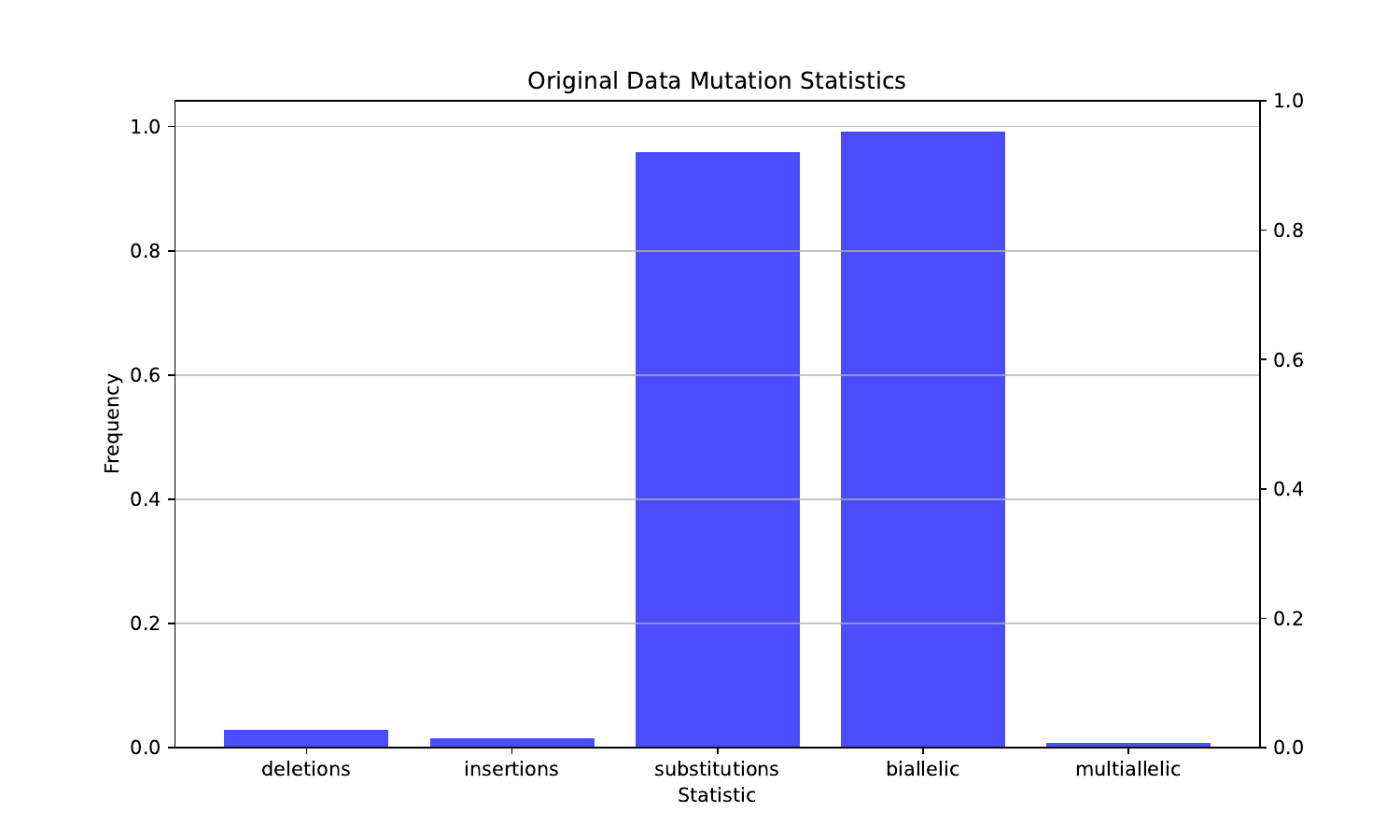}
          \caption{Mutation Statistics of the original 1000GP data.}
    \end{subfigure}
    \hfill
    \begin{subfigure}[t]{0.49\linewidth}
          \centering 
          \includegraphics[width=\linewidth]{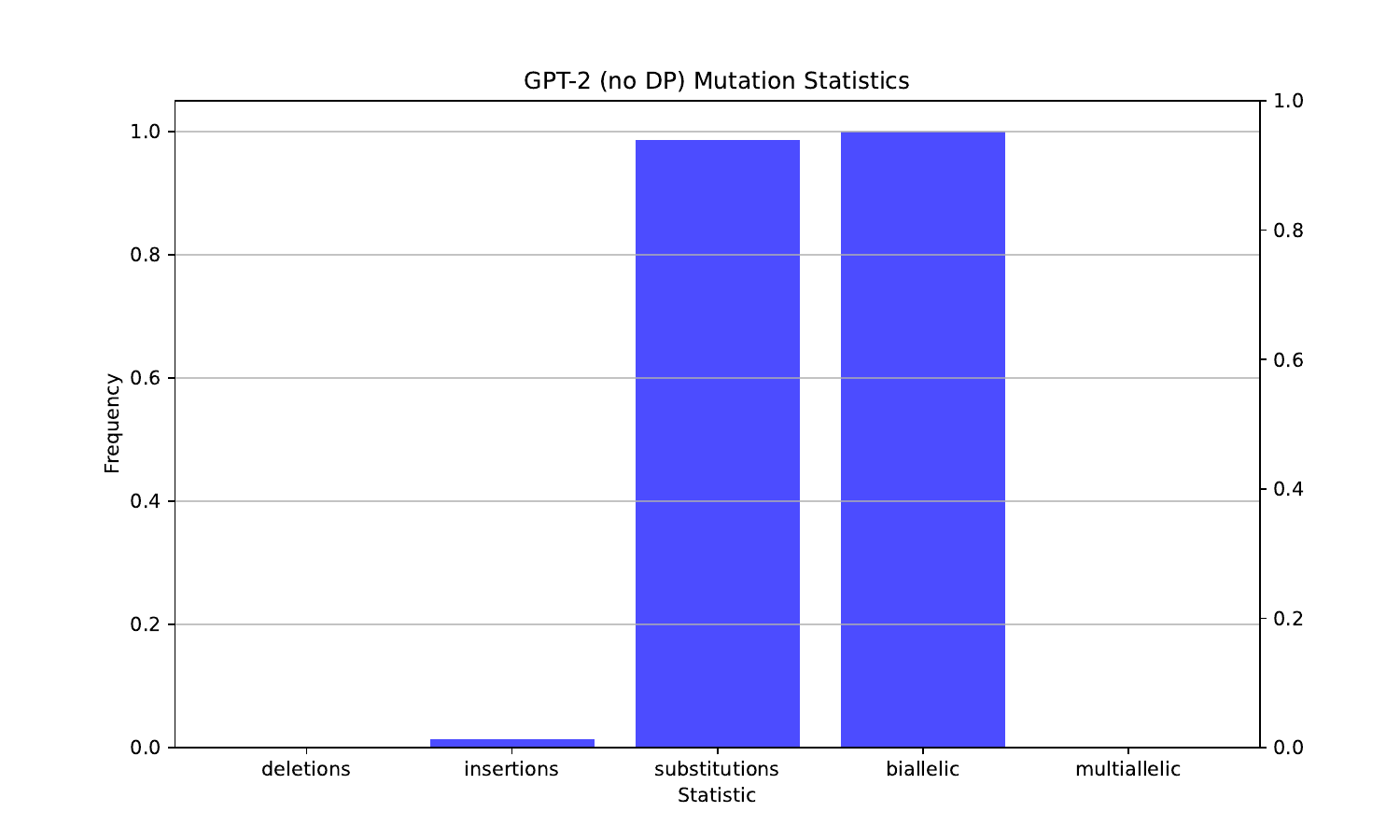}
          \caption{Mutation Statistics of Synthetic data generated via GPT-2.}
    \end{subfigure}
    \par
    \begin{subfigure}[t]{0.49\linewidth}
          \centering
          \includegraphics[width=\linewidth]{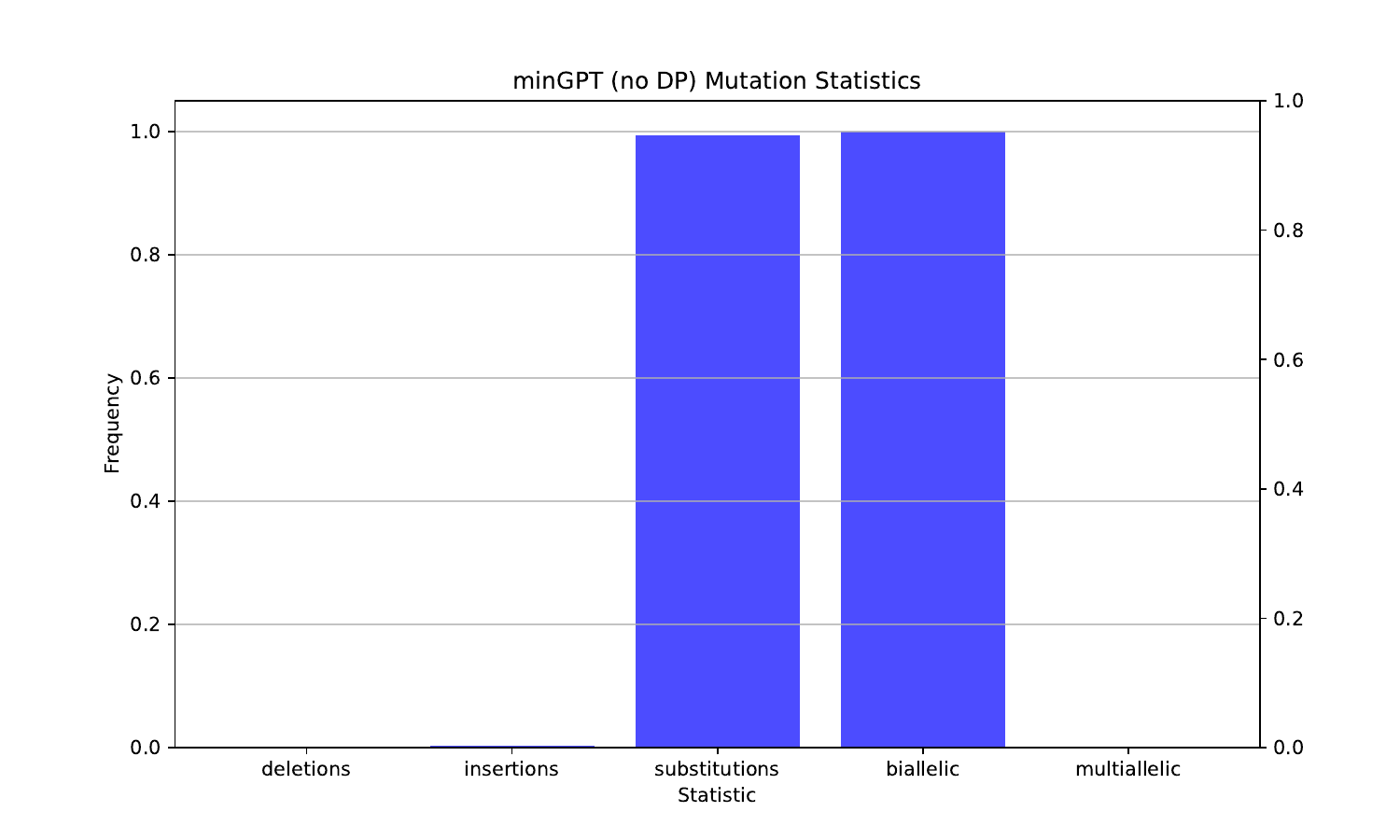}
          \caption{Mutation Statistics of Synthetic data generated via MinGPT (no DP).}
    \end{subfigure}
    \hfill
    \begin{subfigure}[t]{0.49\linewidth}
          \centering 
          \includegraphics[width=\linewidth]{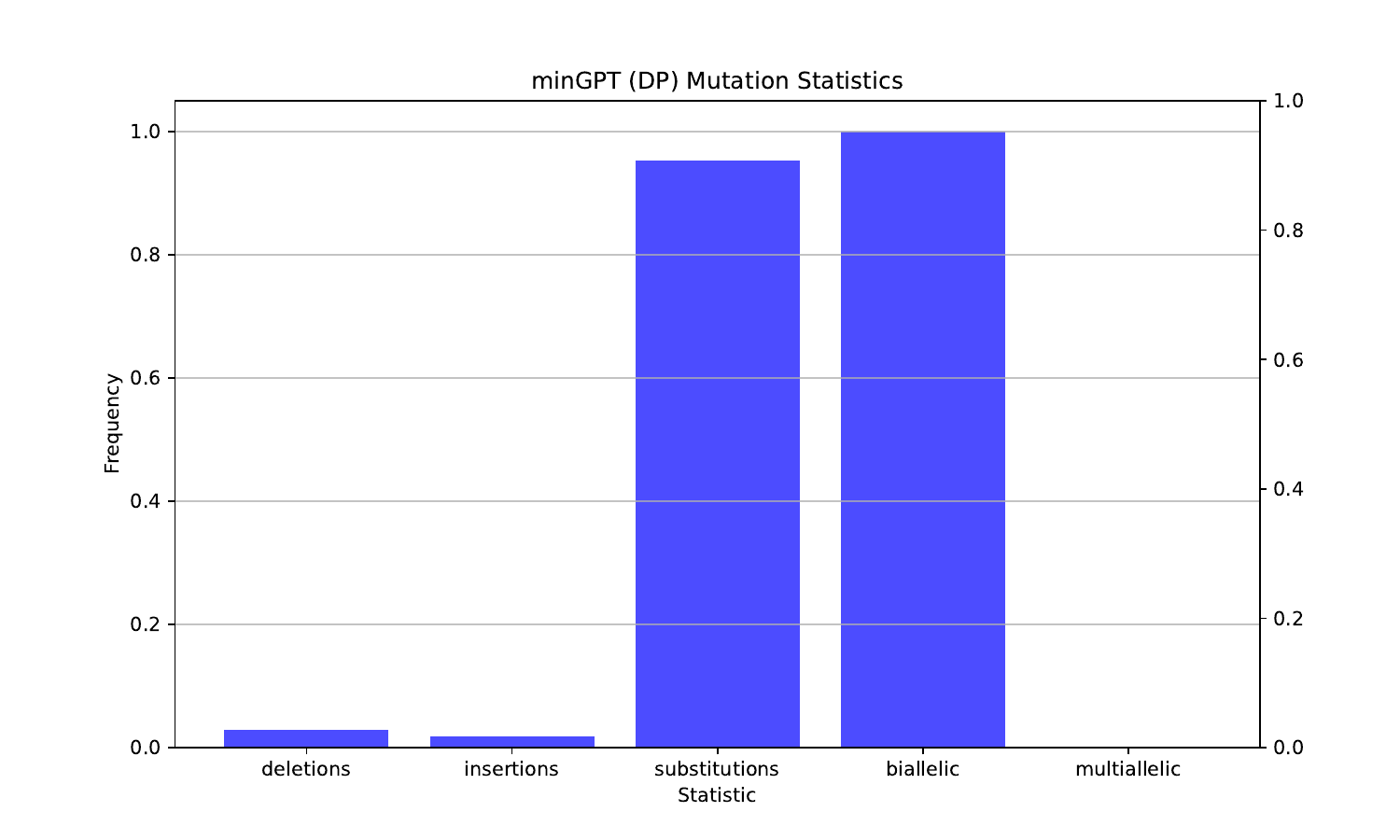}
          \caption{Mutation Statistics of Synthetic data generated via MinGPT (DP $\epsilon=1.0$).}
    \end{subfigure}
    \caption{Comparison of Mutation Statistics metric across synthetic generated cohorts of 50 samples and the original data.} 
    \label{fig:mut-stat-comparison}
\end{figure}

Overall, we found MinGPT-DP to achieve the best utility scores out of all tested models. The model mimics mutation statistics particularly well in terms of mutation type and nature, with frequencies of bi- and multi-allelic and \textit{indels} (insertions and deletions) almost perfectly comparable to the original ones (Figure \ref{fig:mut-stat-comparison}). The finetuned GPT-2 model scores {by far} the highest on mutation memorization, with around to 95\% memorization rate, while both the non-DP and DP trained MinGPT models achieved near-to-none mutation memorization. Furthermore, GPT-2 achieved almost 100\% repetition score on the generated samples, meaning that all samples are generated with the same set of mutations. This caused some overall slight inconveniences, even during the model's privacy evaluation, due to often generating the exact same profiles when prompted with common mutations from the source data, which are realistically shared across most original samples.

\end{document}